\pdfoutput=1 
\documentclass{JINST}

\title{SiPMs characterization and selection for the DUNE far detector photon detection system}

\author{Y. Sun$^a$, J. Maricic$^a$\\
\llap{$^a$}University of Hawaii at Manoa,\\
2505 Correa Rd, High Energy Physics Group, Honolulu, HI, U.S.\\
E-mail: \email{ysun7@hawaii.edu}}

\abstract{The Deep Underground Neutrino Experiment (DUNE) together with 
the Long Baseline Neutrino Facility (LBNF) hosted at the Fermilab will provide a unique, 
world-leading program for the exploration of key questions at the forefront of neutrino physics and astrophysics. 
CP violation in neutrino flavor mixing is one of its most important potential discoveries.
Additionally, the experiment will determine the neutrino mass hierarchy and precisely measure the neutrino
 mixing parameters which may potentially reveal new fundamental symmetries of nature. Moreover, the DUNE is also designed for the 
observation of nucleon decay and supernova burst neutrinos. The photon detection (PD) system in the DUNE far 
detector provides trigger for cosmic backgrounds, enhances supernova burst trigger efficiency and improves the 
energy resolution of the detector. 
The DUNE adopts the technology of liquid argon time projection chamber (LArTPC) that requires the PD sensors, silicon photomultipliers (SiPM), to be 
carefully chosen to not only work properly in LAr temperature, but also meet certain specifications for the life of the experiment. 
A comprehensive testing of SiPMs in cryostat is necessary since the datasheet provided by the manufactures in the market does not cover this temperature regime. 
This paper gives the detailed characterization results of SenSL C-Series 60035 SiPMs, including gain, dark count rate (DCR), 
cross-talk and after-pulse rate. Characteristic studies on SiPMs from other vendors are also discussed in order to avoid any potential problems associated with using a single source. 
Moreover, the results of the ongoing mechanical durability tests are shown for the current candidate, SenSL B/C-Series 60035 SiPMs.}

\keywords{SiPM; Photon Detection; DUNE; Neutrino Oscillation; Supernova; Nucleon Decay; CP Violation; LArTPC}

\begin{document}

\section{DUNE introduction}\label{sec:dune}

Neutrino oscillations are the first evidence beyond the Standard Model. 
The phenomenon of neutrino oscillations was discovered~\cite{SK} more than 40 years after the first detection of neutrinos~\cite{Neutrino_detection_Savannah}. 
Thereafter, a good amount of experiments have been built to measure the relavent oscillation properties. 
After nearly another 2 decades, the unsolved puzzles in the current neutrino oscillation model are CP violation phase and mass ordering.
In addition to answering the unsolved questions and measuring all neutrino oscillation parameters in a single experiment, 
DUNE~\cite{dune} along with LBNF will also provide an opportunity to measure supernova neutrinos and nucleon decay. 

The DUNE experiment uses the most intense neutrino beam in the world from Fermi National Accelerator Laboratory (Fermilab), in Batavia, Illinois, covering a neutrino energy range from 1-10 GeV. 
The primary beamline will provide an initial beam power of 1.2 MW, upgradable to 2.3 MW.
This beamline utilizes the protons from Fermilab Main Injector (MI) to smash on a target to generate a secondary beam primarily composed of charged pions and kaons, which in turn decay to generate the neutrino beam.
The protons extracted from the MI possess energies from 60 to 120 GeV.

The full-scope DUNE far detector is the largest LArTPC built to date located at the 4850 feet level of the Sanford Underground Research Facility (SURF) in Lead, South Dakota.
The DUNE far detector design is scalable and flexible, allowing for a phased approach, with an initial fiducial mass of 10 kt and a final configuration of 40 kt.
The full detector size and its location will enable DUNE to meet the primary scientific goals 
which are to discovery CP violation phase over a large range of $\delta_{CP}$ values, and to significantly advance proton decay lifetime limits as predicted by Grand Unified Theories (GUT). 
The LArTPC technology is able to identify neutrino flavors with a high precision, which offers an excellent sensitivity to proton decay modes with kaons in the final state and provides a unique
sensitivity to electron neutrinos from a core-collapse supernova. 

\section{Photon detector}\label{sec:pd}

The slow ionization-electron drift velocity gives the TPC its 3D imaging capability, but an independent fast signal is required to localize events in time and in space along the drift direction.
The neutrino beam events can be triggered by the machine clock with roughly 10 $\mu$s resolution, which therefore requires no additional trigger.
However, a trigger is required for the non-beam related events, particularly for supernova neutrino and proton decay signals. 

Considering LAr is an excellent scintillating medium, advantage can be taken by triggering on scintillation lights. 
At the existence of 500 V/cm electric field, the light yield of LAr is 24000 photons/MeV.
LAr is highly transparent to its 128 nm scintillation lights with a Rayleigh scattering length and absorption length of 95 cm and $>$200 cm respectively. 
Roughly one third of the photons are prompt ($\tau$=2-6 ns) while two thirds are delayed ($\tau$=1100-1160 ns).
Considering the timing resolution of drift wires is at micro-second level, the prompt photons give instant t$_{0}$ for the event.
In addition to event triggering, detection of the scintillation lights may also be helpful in background rejection.

\begin{figure}[h] 
\centering
\includegraphics[width=.8\textwidth]{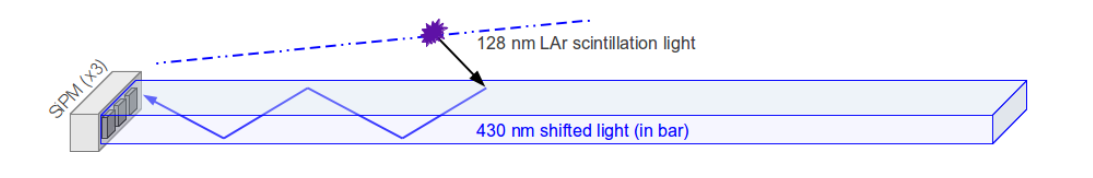}
\caption{Baseline light guide detector design.}
\label{fig:pdbar}
\end{figure}

Supernova events will produce neutrinos down to about 5 MeV.
The PD system is being designed to provide a high-efficiency trigger for supernova events with the lowest possible threshold for a reasonable cost. 
Light collection paddles are employed to reduce the required photon sensor area. 
The most advanced light paddle design uses cast acrylic bars coated with wavelength shifter, and SiPMs are mounted at one end for signal read-out as shown in Fig.~\ref{fig:pdbar}. Details of the PDS design for the DUNE experiment can be found in~\cite{dunepdsdesign}.
SiPMs are selected as PD sensors due to their compact size and low prices. 
Their distinguished properties, especially the very high quantum efficiency compared to the established PMTs, make them ideal replacements of cryogenic PMTs.

\section{SiPM characterization}\label{sec:sipm}

As described in last section, the PD system is designed for non-beam related event timing and supernova neutrinos detection. For the event timing, the requirements for SiPMs are low dark rate for the purpose of low accidental triggering and high gain for the purpose of high SPE efficiency. The supernova neutrino detection requires the SiPMs to be able to detect supernova neutrinos down to about 5 MeV. This in turn requires high gain and low dark rate as described above as well as low cross-talk and after-pulse. The cross-talk and after-pulse will both deteriorate the energy resolution. And the after-pulse will also degenerate the measurement of the prompt to delayed LAr scintillation light ratio. The specific criteria for the gain, dark rate, cross-talk and after-pulse could be derived from the simulation studies which have not been fully developed yet due to changes in the DUNE configuration.

SiPM consists of a Geiger-mode avalanche photodiode (APD) array on common silicon substrate. The applicabilities of APDs in high-energy physics~\cite{SiPM_HEP}, astrophysics~\cite{SiPM_ASP} and medical imaging~\cite{SiPM_MI} have been being studied. 
There are typically 1000 micro-cells (pixels) in a 1 mm$^2$ area parallelly connected in SiPMs. Each pixel bears a similarity of a photodiode and a quenching resistor in series and therefore results in a uniform response. 
The sensitivity regime of SiPMs usually ranges from UV through NIR with capability of single photon resolution. Unlike traditional PMTs, 
SiPMs usually do not require high voltage supply exceeding 100 V. Nonetheless, their gains are comparable to traditional PMTs. Benefiting from the practical advantages of solid-state technology, SiPMs cannot be damaged by stray light. 
Moreover, they are immune to electromagnetic fields and stable with temperature variations. 
The compactness can be scaled up as required. The relative lower cost than the conventional PMTs makes them novel photon detectors and attracts more attention in modern research. 

SenSL's C-Series sensors feature an industry-leading, low DCR and after-pulse rate, combined with high PDE and therefore became our first candidate. Biased at V$_{br}$+5 V, SenSL's C-Series 60035 SiPMs have a 41$\%$ PDE at its peak wavelength of 420 nm. As shown in Fig.~\ref{fig:pdbar}, the 128 nm scintillation lights are wavelength shifted to 430 nm which is around the PDE peak in the spectrum. 
The operational bias voltage in room temperature ranges from 24.65 V (breakdown voltage) to 30 V. 
These newly released C-Series sensors are pin-for-pin compatible with the previous generation B-Series. 
The C-Series 60035 SiPMs have significant improvements in DCR and after-pulse rate. 
The DCR has been reduced from 21.5 MHz to 1.2 MHz at V$_{br}$+2.5 V under 21$^o$C. 
The after-pulse rate has reached 0.2$\%$ at V$_{br}$+2.5 V under 21$^o$C.
The performances of SenSL C-Series SiPMs under room temperature are well characterized and documented by SenSL~\cite{sensl_datasheet}. 
However, the performance characteristics at about LAr temperature have rarely been studied. 
In this section, five SenSL C-Series SiPMs' characteristics such as gain, DCR, cross-talk and after-pulse are examined. Since PDE is not a tunable parameter in the current design, it will not be further discussed in this paper.

\subsection{Experimental setup}\label{sec:Experimental Setup}

\begin{figure}[t] 
\centering
\includegraphics[width=.8\textwidth]{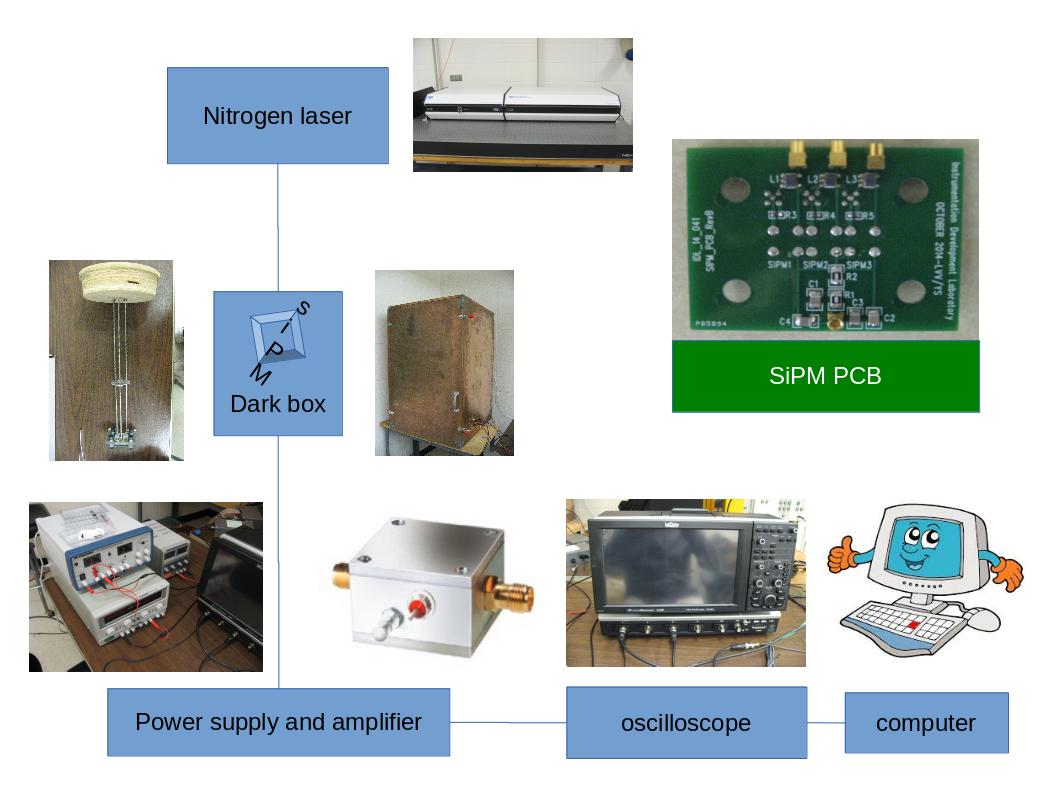}
\caption{The SiPM characterization system setup at UH lab.}
\label{fig:setup}
\end{figure}

\begin{figure}[tbp] 
\centering
\includegraphics[width=.8\textwidth]{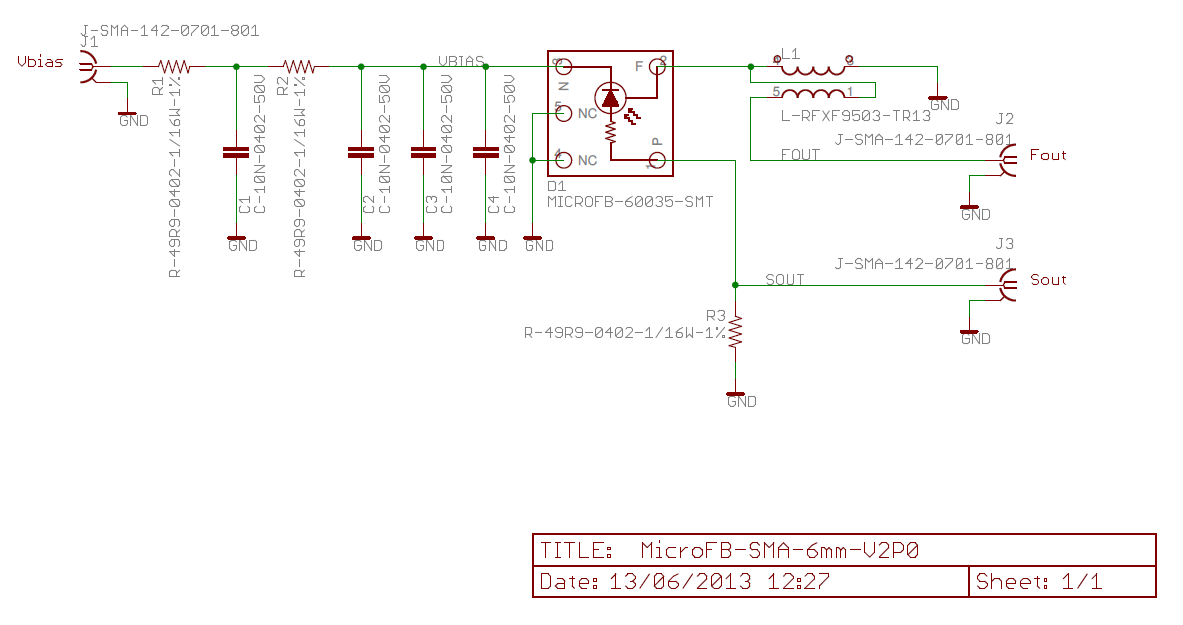}
\caption{The schematic of MicroBF-SMA-6mm provided by SenSL.}
\label{fig:MicroFB_SMA_6mm}
\end{figure}

The experimental setup diagram is shown in Fig.~\ref{fig:setup}. 
The SiPMs are deployed into a cryogenic jar full of LN$_{2}$ (since LN$_{2}$ is only 10 K lower than LAr, no significant performance differences are expected) inside a darkbox which allows BNC cables to power them and read the signals out to the oscilloscope. 
The darkbox prevents any background contamination from ambient light. 
The DAQ computer collects and stores the digital waveforms from the oscilloscope.
Our nitrogen laser system can provide an injection light to the SiPMs through optical fibers.
The PCB is designed by the author based on the MicroBF-SMA-6mm circuit schematics provided by SenSL as shown in Fig.~\ref{fig:MicroFB_SMA_6mm}.
To allow repeat tests for many SiPMs, soldering should be avoided. 
Pogo pins are used to connect the SiPMs to the PCB through mechnical press.
The connection from PCB to the power supply and oscilloscope are through MMCX connectors which provide a mechanical robustness.
The feasibility of this setup has been proved as we have performed many tests in LN$_{2}$.
The amplifiers used in this study are low noise ZFL-1000LN+ from Mini-Circuits~\cite{amp} recommended by SenSL. They are positioned in the darkbox outside the LN$_{2}$ jar.

\subsection{Gain}\label{sec:Gain}

\begin{figure}[h] 
\centering
\includegraphics[width=.7\textwidth]{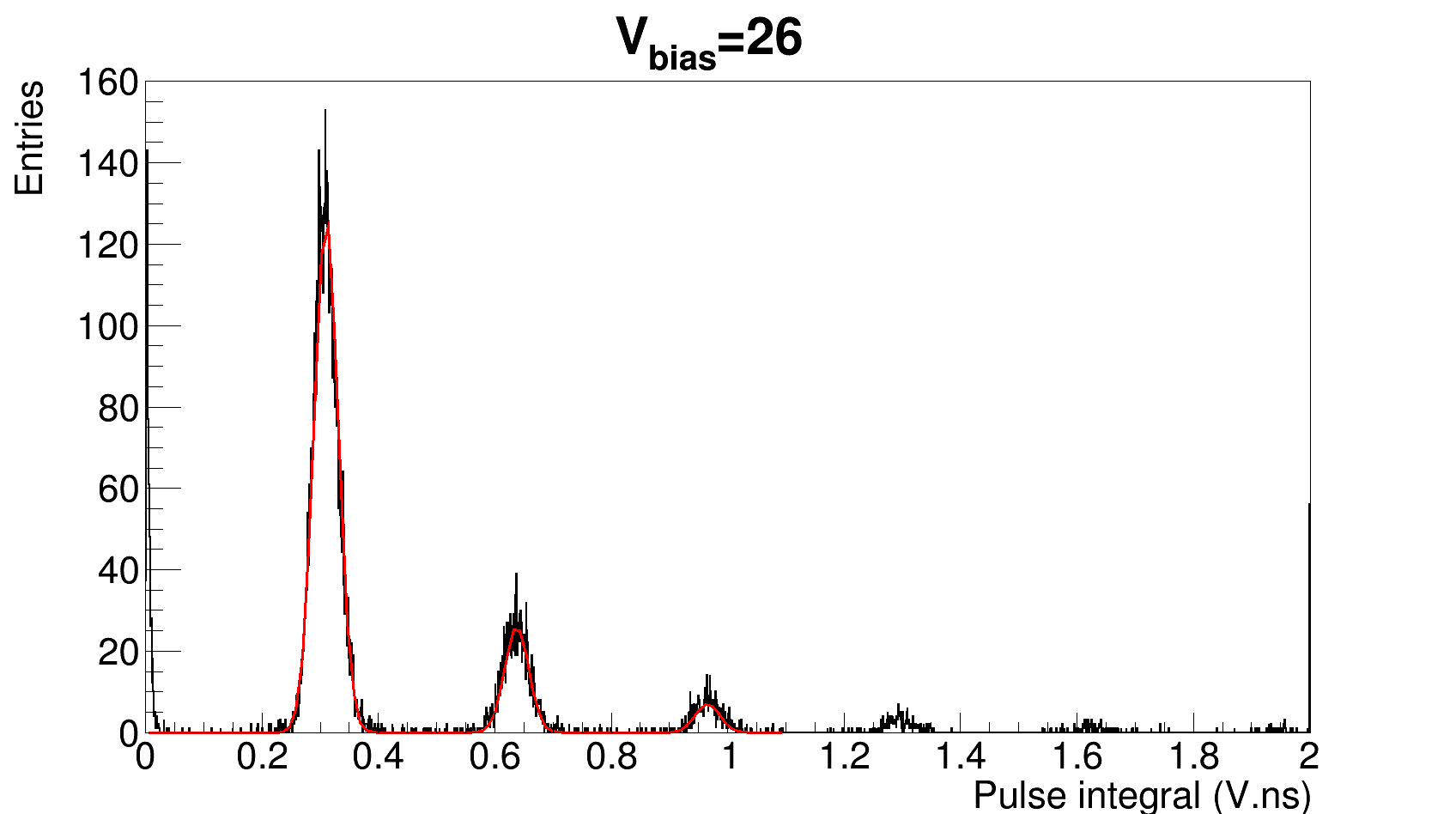}
\caption{SenSL C-Series 60035 SiPM dark pulse integral distribution and its fit for V$_{bias}$ = 26 V.}
\label{fig:pulse_integral_26V}
\end{figure}

\begin{figure}[h] 
\centering
\includegraphics[width=.7\textwidth]{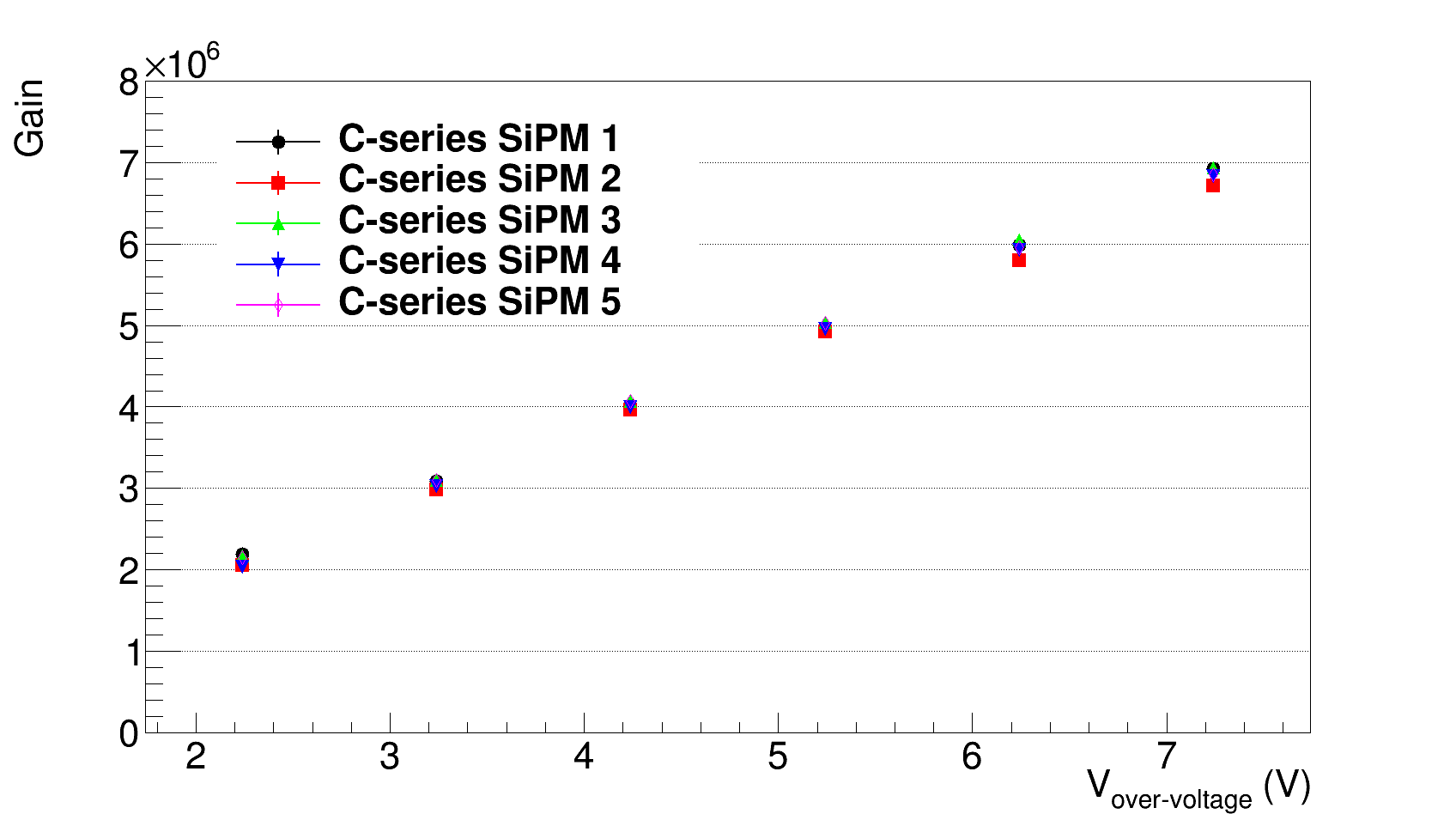}
\caption{Five SenSL C-Series 60035 SiPM gains as a function of over-voltage.}
\label{fig:gain_sipms}
\end{figure}

The SiPM gain is evaluated by using dark pulses. It is given by:
\begin{equation}\label{eq:gain}
G = \frac{\Delta\int U(t) dt}{R\times e\times A},
\end{equation}
where the impedance R = 47.62 $\Omega$, electron charge e = 1.6 $\times$ 10$^{-19}$ C and the amplification A for the amplifiers is calibrated in-situ.
The integral in the numerator can be obtained by integrating the signal pulses. 
Figure~\ref{fig:pulse_integral_26V} shows the integral of dark pulses for V$_{bias}$ = 26 V. 
The NPE peaks are well separated and fitted by~Eq.\ref{eq:fit}.
\begin{equation}\label{eq:fit}
F = \sum_n A_{n} exp [-\frac{(x-nS)^2}{2\sigma^2}]+C,
\end{equation}
where $A_{n}$ is the amplitude of each Gaussian, $S$ is the separation between each Gaussian which gives the value of $\Delta\int U(t) dt$, C is a small overall shift.
The gains behavior as a function of the bias voltage are shown in Fig.~\ref{fig:gain_sipms}. 
A linear dependence in the range of interest can be clearly seen.
The breakdown voltages approximate 21 V as extrapolated at the point where the gain is 0.
Comparasons between SiPMs indicate that the SiPM to SiPM variations are tiny. 

\vspace{0 cm}

\subsection{Dark count rate}\label{sec:Dark Count Rate}
\begin{figure}[h] 
\centering
\includegraphics[width=.7\textwidth]{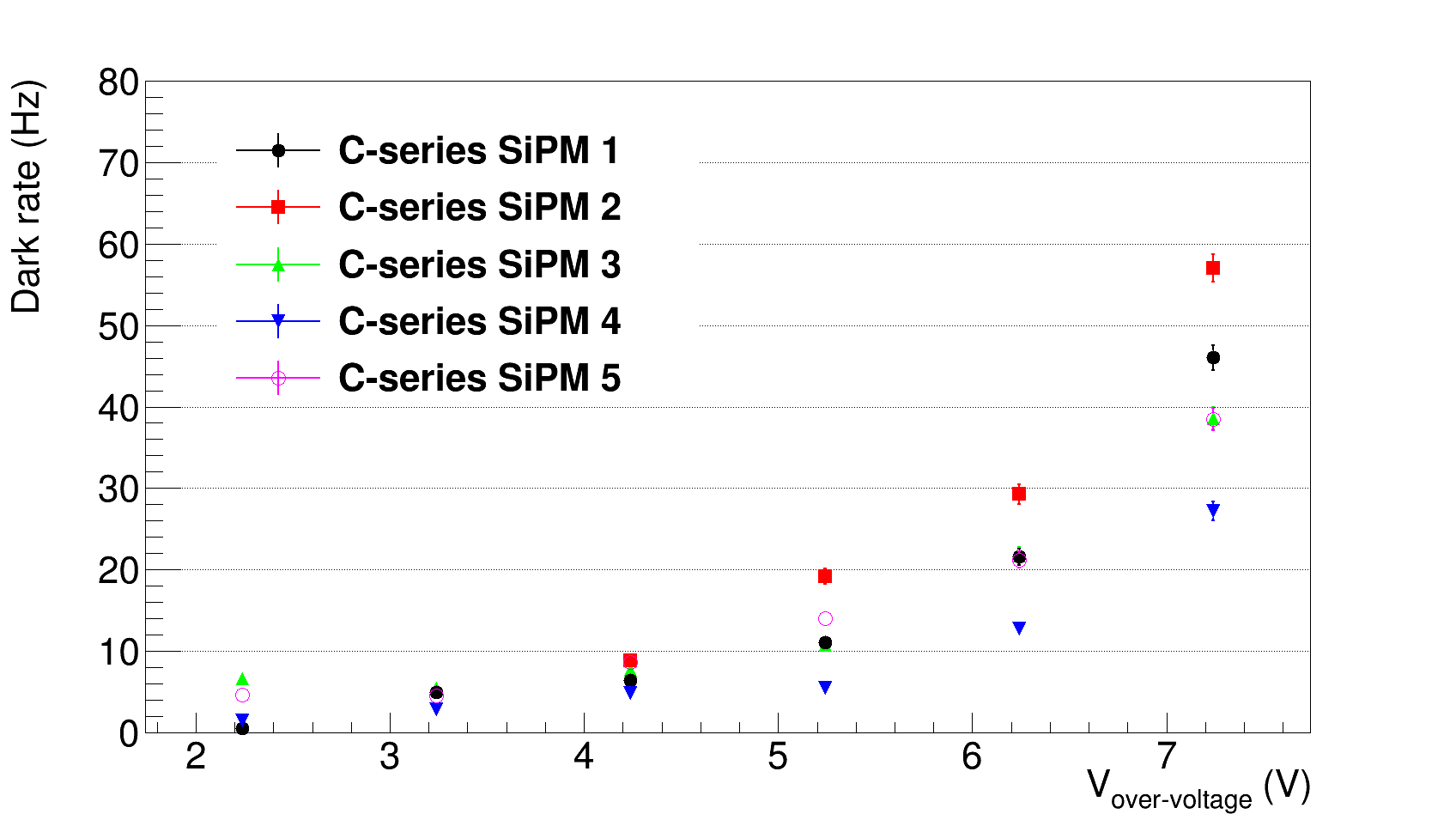}
\caption{Five SenSL C-Series 60035 SiPM DCRs as a function of over-voltage.}
\label{fig:dr_sipms}
\end{figure}

Thermal agitation causing excitations of electrons in the silicon lattice are often referred to as thermal noise or dark noise. 
The frequency of the dark noise is noted as DCR.
In this measurement, 20-second data is collected by random triggering.
DCR increasing exponentially as a function of bias voltage is shown in Fig.~\ref{fig:dr_sipms}.
In general, DCR for all the SenSL C-Series 60035 SiPMs are extremely low in LN$_{2}$. 
The DCR are lower than 5 and 20 Hz at the reference bias voltage V$_{br}$+2.5 V and V$_{br}$+5 V respectively, although there are some variations from SiPM to SiPM.

\subsection{Cross-talk}\label{sec:Cross-talk}
\begin{figure}[h] 
\centering
\includegraphics[width=.7\textwidth]{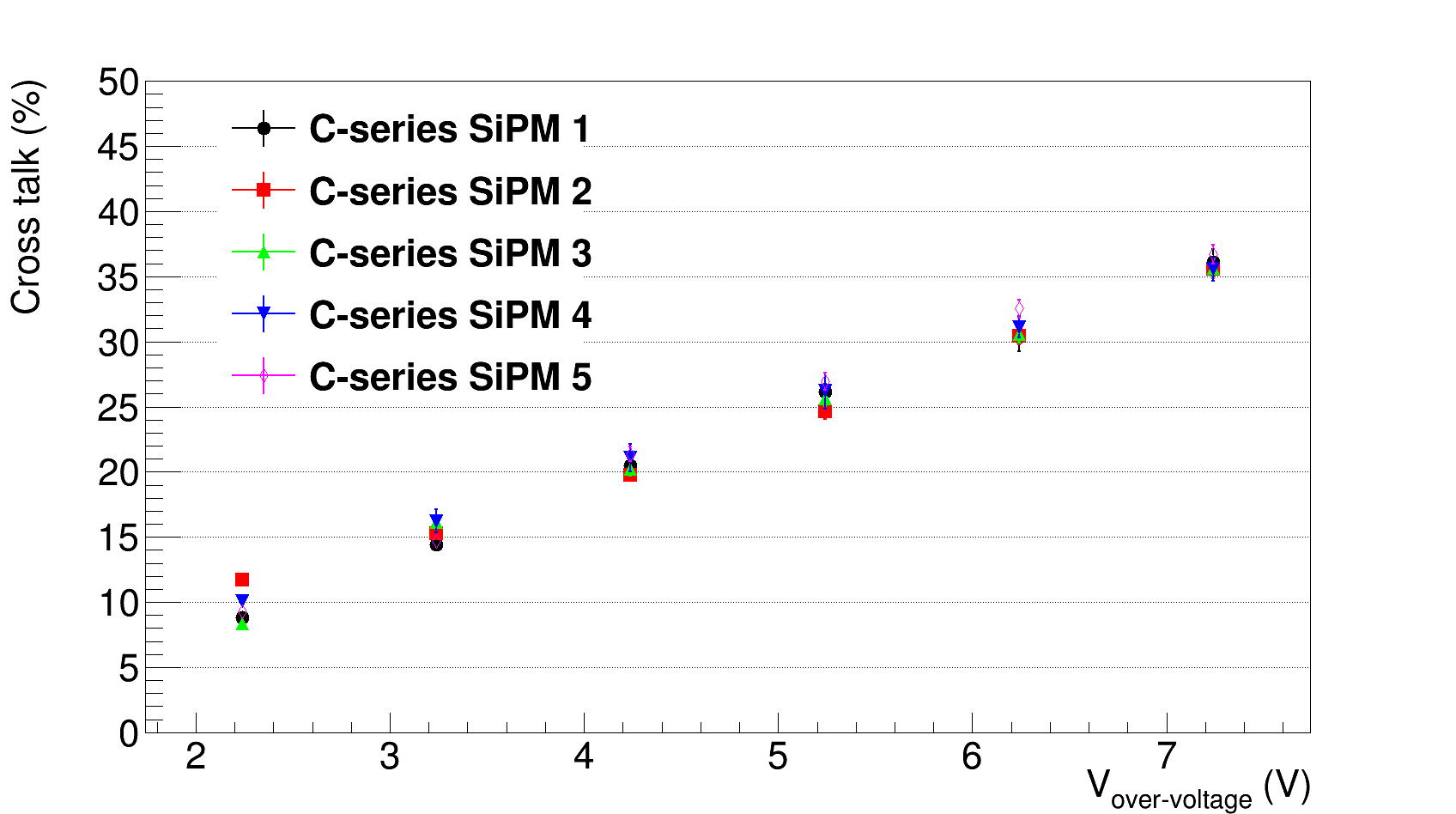}
\caption{Five SenSL C-Series 60035 SiPM cross-talks as a function of over-voltage.}
\label{fig:crosstalk}
\end{figure}

Cross-talk is modelized as photons emitted by the avalanche plasma that reach the neighbour cell diodes and trigger a second avalanche.
Cross-talks degenerate the photon-counting resolution by overcounting the incident photons. 
The method to determine the cross-talk is based on the analysis of signal events generated by dark noise.
Considering the DCR derived above, the probability of two or more
simultaneous thermal excitations for an ideal detector should be negligible.
Only signals with an amplitude corresponding to a single PE should be observed without cross-talk.
However, in reality, cross-talk induced avalanches can occur resulting in higher amplitude pulses as multiple PE peaks can be observed in Fig.~\ref{fig:pulse_integral_26V}. 
By comparing the DCR above a 1 PE threshold with the measured total DCR, the cross-talk probability is estimated.
\begin{equation}\label{eq:crosstalk}
Ct = \frac{DCR_{1.5p}}{DCR_{0.5p}},
\end{equation}
where $DCR_{1.5p}$ and $DCR_{0.5p}$ are the dark noise frequencies measured by setting 
the threshold at 1.5 and 0.5 PE respectively.
The rates $DCR_{1.5p}$ and $DCR_{0.5p}$ are determined automatically by fitting a
spline to the data as shown in Fig.~\ref{fig:pulse_integral_26V}.
Cross-talks also increase linely as a function of bias voltage as shown in Fig.~\ref{fig:crosstalk}.

\subsection{After-pulse}\label{sec:After-pulse}
After-pulses are generated when electrons
produced in an avalanche are trapped and released
again after some delay which can last from nanoseconds
up to several microseconds. These signals cannot be separated from
genuine, photon-induced signals and thus deteriorate
the photon-counting resolution. The measurement of the after-pulse probability is based on the analysis of a weak injected light. 
The trigger is applied on the injected pulse and after-pulse searching lasts for 20 $\mu s$ after the main pulse.

\begin{figure}[t] 
\centering
\includegraphics[width=.7\textwidth]{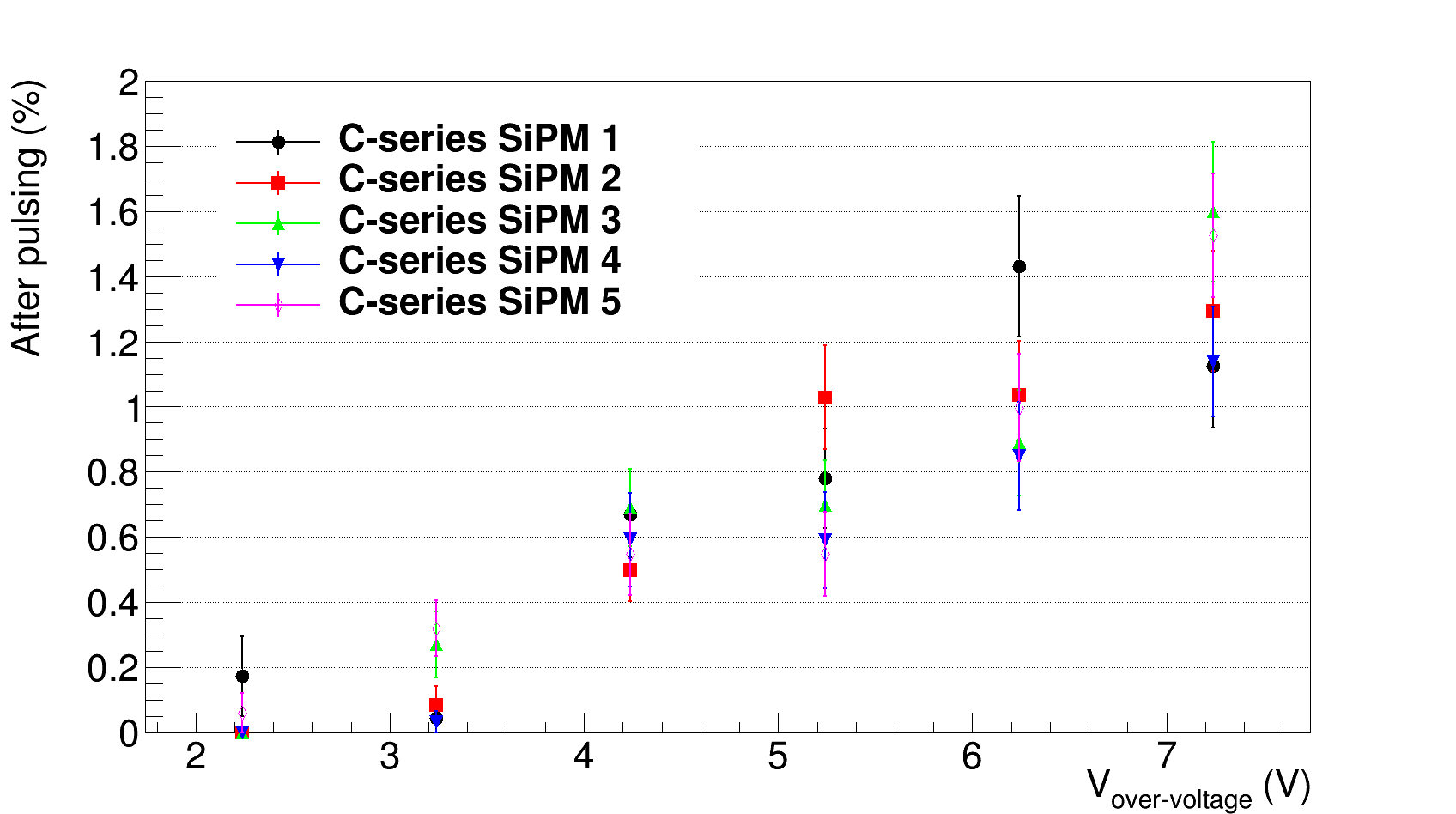}
\caption{Five SenSL C-Series 60035 SiPM after-pulse rates as a function of over-voltage.}
\label{fig:ap}
\end{figure}

\begin{figure}[h] 
\centering
\includegraphics[width=.7\textwidth]{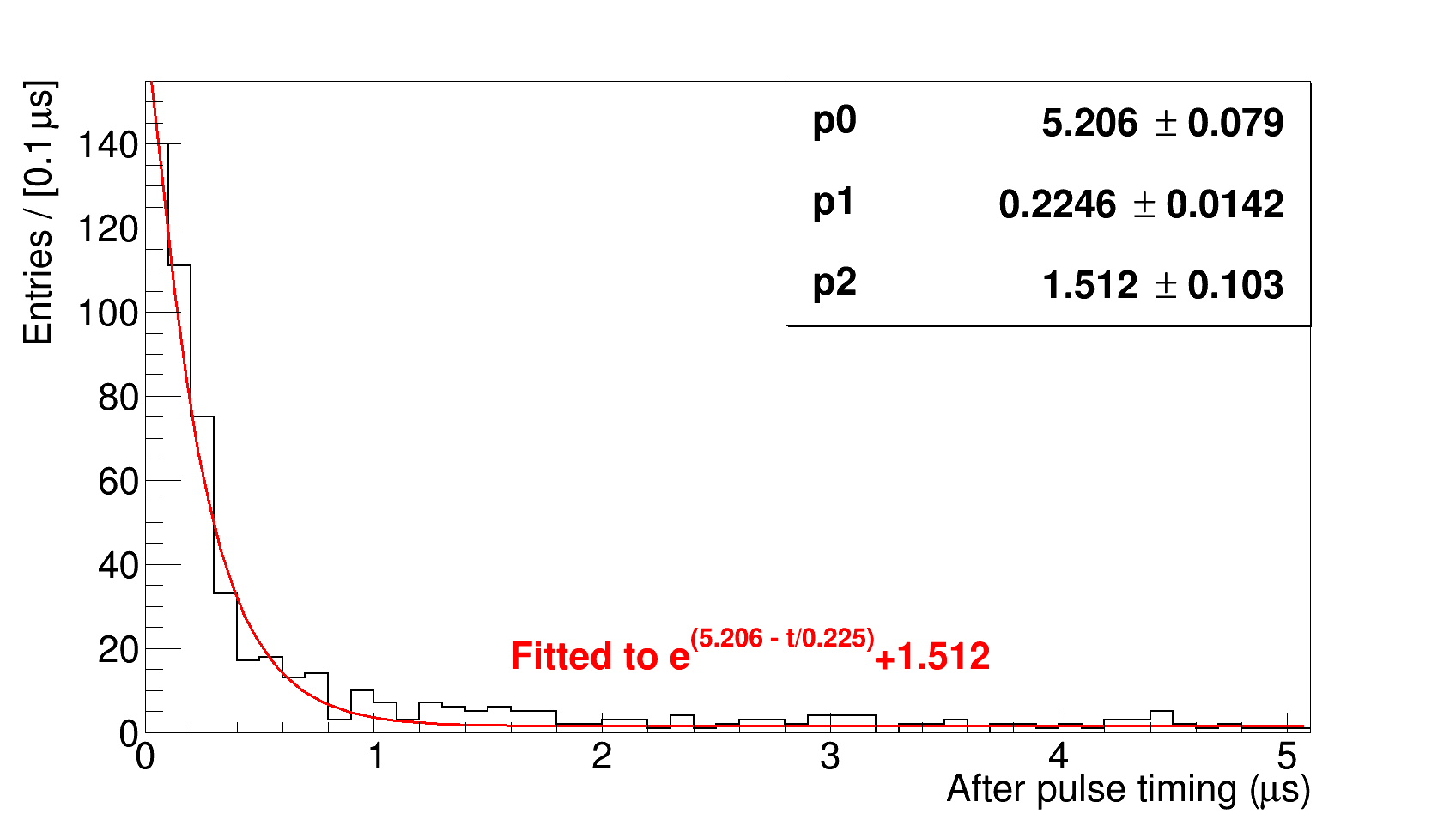}
\caption{After-pulse timing fitted to an exponential plus a constant as shown in the figure.}
\label{fig:ap_timing}
\end{figure}

The characterization results of after-pulse are summarized in Fig.~\ref{fig:ap}. 
The datasheet gives 0.2$\%$ after-pulse probability at V$_{br}$+2.5 V under 21$^o$C. 
The after-pulse rate in LN$_{2}$ agrees with the probability in the room temperature.
As the bias voltage increases, the after-pulse rate goes up but they are still less than 1.6$\%$ even biasd at V$_{bias}$ = 28 V. 
In a word, the after-pulse rate is extremely low for SenSL C-Series SiPMs.
The after-pulse timing is summarized in Fig.~\ref{fig:ap_timing}. The fit to an exponential plus a constant gives a time constant of 224.6 ns. 

\section{SiPM selections and durability tests}\label{sec:selection_durability}
To prevent any potential problem from using a sole source of SiPMs, selection of at least two models from two different vendors is needed. 
We have tested SiPMs from several manufactures, such as Hamamatsu, NDL, AdvanSiD and KETEK. 
On the other hand, mechanical damages have been observed on SenSL SiPMs over cryogenic cycles. 
In order to understand the mechanism of the damages as well as plan for the best strategy to the inevitable cyclings, a mechanical test plan had been made and being executed.

\subsection{SiPM selections}\label{sec:selection}

\begin{figure}[h] 
\centering
\includegraphics[width=.9\textwidth]{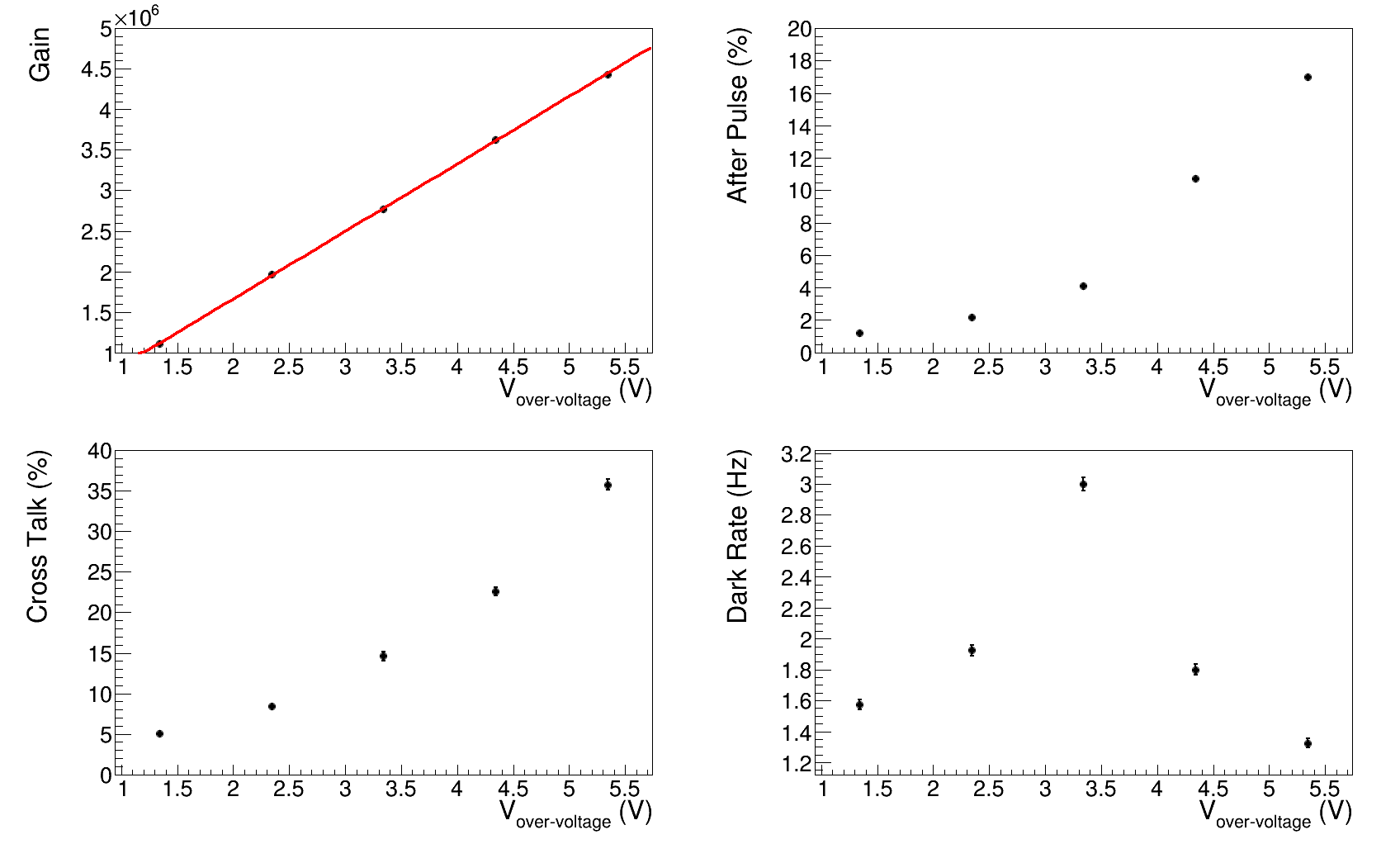}
\caption{Gain, after-pulse rate, cross-talk and DCR as a function of over-voltage for a Hamamatsu SIL-LCT SiPM.}
\label{fig:hamamatsu}
\end{figure}

NDL 11-2222B-S and AdvanSiD NUV SiPMs failed cryogenic tests. Since they both function properly at room temperature even after cryogenic cycles, the failures are probably due to the intrinsic defects with the silicon wafers working at LN$_{2}$ temperature. 
KETEK PM6660 has a reduced operative range in LN$_{2}$ and 2 orders of magnitude more after-pulse rate than SenSL C-Series 60035 SiPMs.
Hamamatsu SIL (silicone resin)-STD (standard) and TFC (thin film coating)-STD SiPMs have extremely high after-pulse rate similar to KETEK products.
Hamamatsu SIL-LCT (low cross-talk) and TFC-LCT SiPMs have a reduced after-pulse rate but still approximately 10 times more than SenSL products as shown in the upper-right of Fig.~\ref{fig:hamamatsu}. The characteristics of SenSL C-Series and HAMAMATSU SIL-LCT SiPMs at 2.5 V over-voltage are listed in Tab.~\ref{tab:comparison}

\begin{table}[h]
\label{tab:comparison}
\smallskip
\centering
\begin{tabular}{|ccccc|}
\hline
&Gain&Dark rate (Hz)&Cross-talk&After-pulse\\
\hline
SenSL C-Series SiPMs&2.4$\times$10$^6$&3 (0.083 Hz/mm$^2$)&12$\%$&0.2$\%$\\
\hline
HAMAMATSU SIL-LCT SiPMs&2.1$\times$10$^6$&2 (0.222 Hz/mm$^2$)&10$\%$&2.5$\%$\\
\hline
\end{tabular}
\caption{Characteristics comparison of SenSL C-Series and HAMAMATSU SIL-LCT SiPMs at 2.5 V over-voltage.}
\end{table}

\subsection{SiPM mechanical durability tests}\label{sec:mech}

\begin{figure}[h] 
\centering
\includegraphics[width=.9\textwidth]{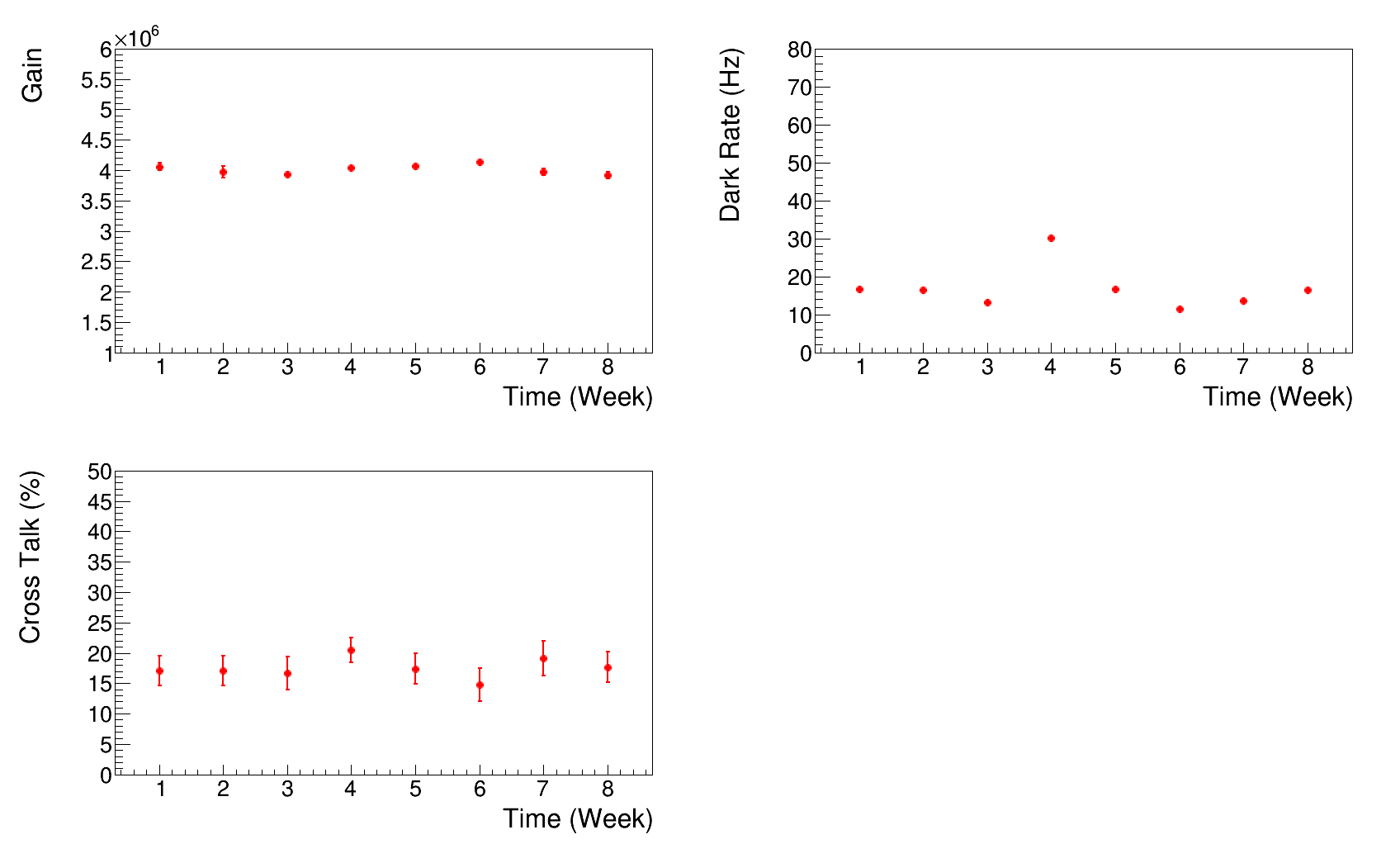}
\caption{Gain, DCR and cross-talk as a function time for one of SenSL B-Series 60035 SiPMs in group A.}
\label{fig:mechanical_tests}
\end{figure}

The tests start with 9 brand new SenSL B-series 60035 SiPMs.
These 9 SiPMs are categorized into 3 groups, namely A, B and C.
Groups A and B are cycled with 7 days soak in LN$_{2}$ while group C has a 2-month long term cycle. 
At the end of each weekly cycle, group A is taken out of the LN$_{2}$ and warms up in a plastic bag which provides a pure nitrogen recovery environment.
Group B warms up naturally in the lab exposed to the air.
After each cycle, all 3 SiPMs in that particular group are visually inspected under microscope and characteristically checked on gain, DCR and cross-talk.
Figure~\ref{fig:mechanical_tests} depicts the characteristics evolution over a time period of 2 months for one SiPM in group A. There is no obvious degradation or change of any characteristic.
A similar test being conducted at Louisiana State University using a combination of both B and C-Series SiPMs also shows no change in DCR or gain detected to date~\cite{cycling}.

\subsection{SiPM aging tests}\label{sec:aging}
  
Besides mechanical durability tests, an aging test~\cite{aging} in Indiana University meant to find out the characteristic behavior under long term cryogenic exposure.
Six SiPMs are kept continuously immersed in LN$_{2}$ since March 9, 2015. 
Three of them are biased at nominal 24.5 V bias voltage (over-voltage $\approx$ 3.5 V) as a control group while the other 3 are biased at 30.5 V bias voltage (over-voltage $\approx$ 9.5 V) as an accelerated aging group.
Measurements have been taken every few days.
After 10 measurements, SiPM properties between two groups are very consistent.
No significant changes in DCR or gain have been observed.

\section{Conclusion}\label{Conclusion}

The SenSL C-Series 60035 SiPM characterization results are consistent with datasheet considering cryogenic temperature. 
The characteristics of the selected SenSL C-Series 60035 SiPMs are suitable for DUNE PD system.
Until the specific requirements for DUNE PD system are fully developed, we will simply keep seeking alternatives. Once we know the maximum tolerance values of dark rate, cross-talk and after-pulse rate, alternatives can be chosen.
Currently, no product in market has competitive devices to SenSL.
However, technology in this area is developping dramatically. 
The manufacturers mentioned in Sec~\ref{sec:selection_durability}. have promised to provide their newest technology to us for the first test.
Thermal cycling tests and aging tests are ongoing. The up-to-date results show no functional damages or aging effects.

\acknowledgments

We gratefully acknowledge the cooperation of the DUNE collaboration in providing their test results.
We also wish to acknowledge the support of the Department of Energy for funding our lab construction and operation.
We thank Marc Rosen, Vihtori Virta and Ziru Sang for their technical supports.


\begin{thebibliography}{9}

\bibitem{SK}
\textbf{SK} Collaboration, Y.~Fukuda et al., \emph{Evidence for oscillation of atmospheric neutrinos}, \emph{Phys. Rev. Lett.} {\bf 81} (1998) {1562--1567}.

\bibitem{Neutrino_detection_Savannah}
C.~L. Cowan, F.~Reines, F.~B. Harrison, H.~W. Kruse, and A.~D. McGuire,
\emph{Detection of the Free Neutrino: a Confirmation,},
\emph{Science} \textbf{124} (1956) 103--104.

\bibitem{dune}
\textbf{DUNE} Collaboration, C. Adams et al., \emph{Scientific Opportunities with the Long-Baseline Neutrino Experiment},
\href{http://xxx.lanl.gov/abs/1307.7335}{\it{Arxiv: 1307.7335}}.

\bibitem{dunepdsdesign}
D. Whittington,
\emph{Photon Detection System Designs for the Deep Underground Neutrino Experiment},
\href{http://arxiv.org/abs/1511.06345}{\it{Arxiv: 1511.06345}}.

\bibitem{SiPM_HEP}
V. Andreev et al.,
\emph{A high-granularity plastic scintillator tile hadronic calorimeter with APD readout for a linear collider detector},
\emph{Nucl. Instr. and Meth. A} \textbf{564} (2006) 144--154.

\bibitem{SiPM_ASP}
A. Biland et al.,
\emph{First detection of air shower Cherenkov light by Geigermode-Avalanche Photodiodes},
\emph{Nucl. Inst. and Meth. A} \textbf{595} (2008) 165--168.

\bibitem{SiPM_MI}
S. Moehrs et al.,
\emph{A detector head design for small-animal PET with silicon photomultipliers (SiPM)},
\emph{Phys. Med. Biol.} \textbf{51} (2006) 1113--1127.


\bibitem{sensl_datasheet}
SenSL,
\href{http://sensl.com/downloads/ds/DS-MicroCseries.pdf}
{\emph{C-Series Low Noise, Fast, Blue-Sensitive Silicon Photomultipliers DATASHEET}}


\bibitem{amp}
Mini-Circuits,
\href{http://www.minicircuits.com/pdfs/ZFL-1000LN+.pdf}
{\emph{Low Noise Amplifier ZFL-1000LN+}}


\bibitem{cycling}
Thomas Kutter,
\emph{Silicon Photomultiplier Requirements and Testing},
\emph{LBNE internal document} \textbf{10881}

\bibitem{aging}
S. Mufson et al.,
\emph{Recent Progress in Bar Technology/SiPM Aging Studies at Indiana University},
\emph{LBNE internal document} \textbf{10817}


\end{thebibliography}
\end{document}